# Analysis of Gradient based Algorithm for Signal reconstruction in the Presence of Noise

Slavoljub Jokić, Ljindita Niković, Jelena Kadović
Faculty of Electrical Engineering
University of Montenegro
Podgorica, Montenegro
slavoljub.jokic@gmail.com, linditanikaj@gmail.com, kadovicjelena7@gmail.com

*Abstract*— **Common problem in signal processing is reconstruction of the missing signal samples. Missing samples can occur by intentionally omitting signal coefficients to reduce memory requirements, or to speed up the transmission process. Also, noisy signal coefficients can be considered as missing ones, since they have wrong values due to the noise. The reconstruction of these coefficients is demanding task, considered within the Compressive sensing area. Signal with large number of missing samples can be recovered, if certain conditions are satisfied. There is a number of algorithms used for signal reconstruction. In this paper we have analyzed the performance of iterative gradient-based algorithm for sparse signal reconstruction. The parameters influence on the optimal performances of this algorithm is tested. Two cases are observed: non-noisy and noisy signal case. The theory is proved on examples.**

## I. Introduction

The recent studies have shown that, under certain conditions, successful signal reconstruction is possible, using significantly smaller number of samples than required by the conventional sampling theorem. A great attention has been dedicated to this technique [1]-[3], called Compressive Sensing (CS), because it allows us to acquire and store smaller amount of samples (data) than usual. The CS combines sampling and compression into one step by acquiring small number of signal samples that contain maximum information about the signal. This approach can be efficiently applied to signals which can be represented with small number of non-zero coefficients in a certain domain [3]. Also, the measurement procedure should provide reconstruction from small number of acquired samples, which means incoherence property should be satisfied. Moreover, it is shown that incoherence is assured by randomly acquiring signal samples.

The reconstruction from small set of acquired samples is enabled by using optimization algorithms. There is a large number of optimization algorithms, based on different norm minimizations. Commonly used are algorithms based on convex optimization using $l_1$ norm minimization [4]-[7], but there is an intensive growth of the optimization techniques. A large number of algorithms is based on approximate solutions based on thresholding or greedy methods [8]-[14]. In this paper, we have analyzed the performance of the iterative algorithm for reconstruction of sparse signals, based on gradient calculation and signal concentration as a measure of sparsity [6],[7]. When the iterations approach the optimal point, gradient value oscillates around the true value. We have used the norm one in the examples. The samples are randomly selected from the time domain of the signal. The reconstruction efficiency is observed in the case of non-noisy and noisy signals, and the performance is analyzed in terms of reconstruction accuracy, using the signal to noise ration and mean absolute error as quality metrics.

The paper is structured as follows: The basic theory behind CS is given in the Section II. In Section III adaptive gradient algorithm is described, while the experimental results are in the Section IV. Conclusion is given in the Section V.

## II. Compressive Sensing basics

The CS relies on two main principles: sparsity and incoherence. The incoherence states that samples must be acquired randomly over the whole domain so that we acquire all the useful information of the signal. There are several ways used to reconstruct sparse signals and they are meanly divided in two groups: the group based on the gradient approach and the matching pursuit approach. For the reconstruction we use the measure of sparsity of a signal as a minimization function. This measure represents the number of non-zero coefficients in a signal and can mathematically be modeled as different norms. Norm zero presents the sum of the transformations absolute value on power zero, thus this norm counts the number of non-zero coefficients. The zero norm is very sensitive to disturbance. Disturbance can cause the zero coefficients (which due to this disturbance have a greater value then zero) to be equal to one and counted as non-zero coefficients**.** Therefore, we use other norms that are less sensitive to disturbance like norm one ($l_1$), norm two ($l_2$), norms in between $l_1$ and $l_2$ and greater norms to avoid this problem. Based on the ratio of different norms we can evaluate the concentration measure of a transform used in the gradient approach to directly reconstruct the missing samples:

$$M\left[T\left[s(n)\right]\right]=\frac{1}{N}\sum_i |S(i)|^{1/p} \qquad (1)$$

where $s(n)$ is the signal that is sparse in transformation domain $S(i)=T[s(n)]$, $N$ the number of samples of $S(i)$ and $1\leq p\leq\infty$ is the norm employed.

In this paper we will focus on a gradient based algorithm [6],[7] for which we will propose different parameters later on, first we will analyze the gradient based algorithm. This algorithm performs a direct search over all missing samples of the signal. If the values of every missing sample are located in the range of −A to A then for every missing sample the algorithm performs a search over all possible values in the given range by taking any given step. Larger steps are taken in the first few ruff approximations. As we get close to the true value of the missing samples we go about lowering the step size to achieve a desired precision. This adaptive variable step implemented in the algorithm enables the reconstruction to be done in a significantly smaller number of iterations and with an acceptable precision. A brief presentation of the algorithm used for reconstruction: Assume $s(n)$ is a discrete signal and $T[s(n)]$ its transformation in which $s(n)$ is sparse.

Step 0:

$$x^{(0)}(n) = \begin{cases} s(n), & \text{for available samples} \\ 0, & \text{for missing samples} \end{cases} \quad (2)$$

In the first step we form a vector containing the values of the sampled signal and the zero coefficients of the same signal [7].

Step 1:

$$x_1^{(i)}(n) = \begin{cases} x^{(i)}(n) + d & \text{for } n = n_k \\ x^{(i)}(n) & \text{for } n \neq n_k \end{cases} \quad (3)$$

$$x_2^{(i)}(n) = \begin{cases} x^{(i)}(n) - d & \text{for } n = n_k \\ x^{(i)}(n) & \text{for } n \neq n_k \end{cases} \quad (4)$$

$n=n_k$ index of the missing sample. In the second step each missing sample is increased or decreased for the value of the step $d$.

Step 2:

$$e(n_k) = \frac{M\left[T\left[x_1^{(i)}(n)\right]\right] - M\left[T\left[x_2^{(i)}(n)\right]\right]}{2d} \quad (5)$$

Estimate the signal differential measure for each missing sample, $e(n_k)=0$ for available samples.

Step 3:
Create a gradient vector $E$ containing the values of the estimated differential measures from Step 2.

Step 4:

$$x^{(i+1)}(n) = x^{(i)}(n) - \mu E(n) \quad (6)$$

Based on gradient, vector **E** corrects the values of missing samples. Parameter μ is a constant parameter that determines the precision of the reconstructed samples and their speed of convergence towards the correct values. As we can see the key parameters of which the performance of the algorithm depends on are: the step size $d$, the constant $\mu$ and the norm $p$ used in

Step 2 for calculating the concentration measure of the transform by (1).

### III. ANALYSIS THROUGH EXAMPLES

*Case 1*: Let us consider the signal in the following form:
$$s = 3\sin(2\pi 10t/N) + \sin(2\pi 15t/N)$$

$N$=128 is the total number of samples. We assume that only half of the samples are missing and gradually increase the number of missing samples until the reconstruction is no longer possible. Note that the position of the missing samples is known. First we will analyze the case in which parameters $d$, $\mu$ and norm $p$ are all constant.

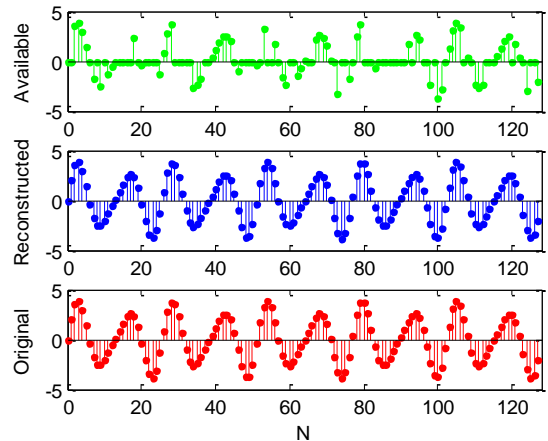

Fig. 1. The original, available and reconstructed signal in case of 64 (half) missing samples, d=0.5, μ=1 and p=1 are constant for each iteration

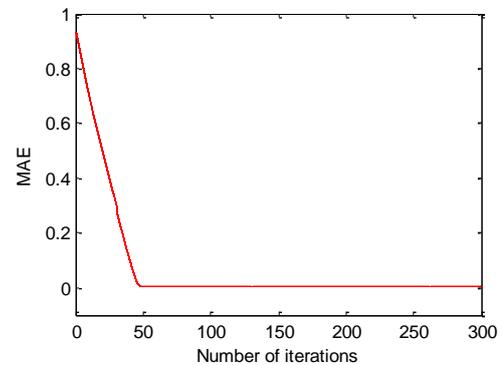

Fig. 2. Mean absolute error for signal presented in Fig.1

In this case we achieve the precision of $10^{-2}$ after 50 iterations. For constant algorithm parameters MAE (Fig. 2) is small but still notable and cannot be improved by increasing number of iterations. Lower values of $d$ and μ produce lower MAE but simulation needs more iterations to obtain satisfactory results.

As we will see in the next case (Fig. 3) by implementing adaptive step size $d$ and adaptive μ the reconstruction can be done faster and with the precision of up to $10^{-15}$.

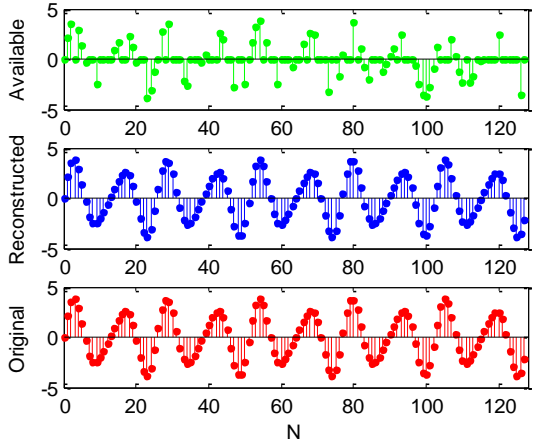

Fig. 3. Original, available and reconstructed signal in case of 64 (half) missing samples, adaptive size of d and μ, p=1

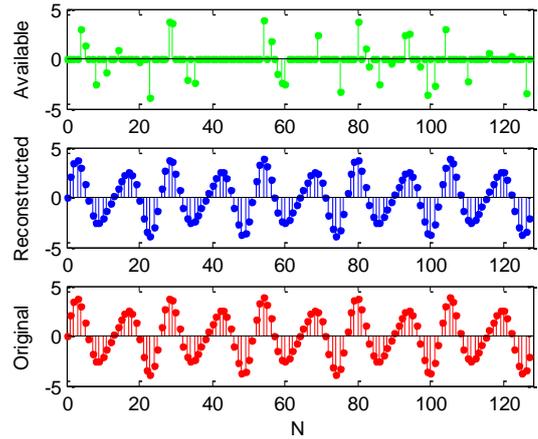

Figure 5: Original, available and reconstructed signal in case 94 (73%) missing samples, adaptive size of d and μ, p=1

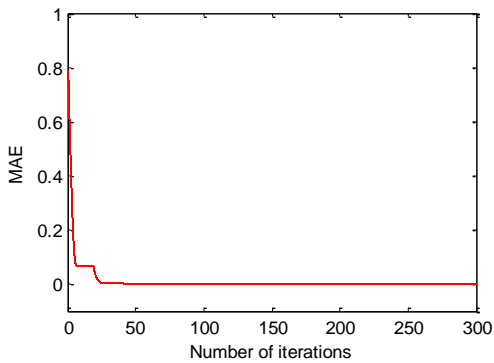

Fig. 4. The mean absolute error for signal presented in Fig.3

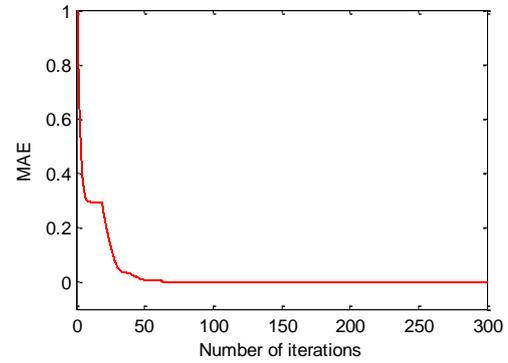

Fig. 6. The main absolute error for signal presented in Fig.5

In this case the starting values of $d$ and $\mu$ are 5 and 10 respectively. We decrease their values by 10 after 20 iteration so that the MAE (Fig.4) falls faster than in previous case (Fig.2). The reconstruction is done with the precision of $10^{-15}$ using the adaptive step sizes even when more than half of the signal samples are missing.

*Case 2:* The same signal form is observed. We will present the case in which 94 for 128 samples are missing (Fig. 5). In this case, we consider the variable sizes of $d$ and $\mu$. After 20 iterations step $d$ and parameter $\mu$ will be decreased by 10 in order to achieve a lower mean absolute error (MAE) comparing to previous cases. The starting values of $d$ is 10 and of $\mu$ is 20. They are decreased by 10 after 20 iterations. As we can see the MAE reaches $10^{-15}$ even for case of 73% unavailable samples.

Simulations show that we get the best results when parameter $\mu$ is twice bigger then step size($d$): $\mu=2d$.
In Table 1, we have numerically shown that (7) must be satisfied in order to achieve the greatest possible precision in the smallest number of iterations. Analysis are done for three different cases of missing samples, while $N=128$ is the total number of samples, norm $p=1$ and value for which $d$ and $\mu$ are decreased is 10 after 20 iterations.

| d | μ | Missing samples | $MAE_{MIN}$ | Number of iterations for $MAE_{MIN}$ |
|---|---|---|---|---|
| 20 | 40 | 64 | $5.86*10^{-15}$ | 283 |
| 10 | 20 | 64 | $5.657*10^{-15}$ | 282 |
| 10 | 10 | 64 | $6.883*10^{-15}$ | 285 |
| 20 | 20 | 64 | $9.575*10^{-15}$ | 286 |
| 20 | 40 | 94 | $7.73*10^{-15}$ | 303 |
| 20 | 20 | 94 | $1.773*10^{-3}$ | 145 |
| 20 | 40 | 35 | $2.04*10^{-15}$ | 287 |
| 20 | 20 | 35 | $4.794*10^{-15}$ | 281 |

Table 1. Suitable values of d and μ

*Case 3:* Now we will analyze the case in which the same signal is corrupted with noise. The parameters for this case are the same as in *Case 2* in order to compare performance of the algorithm in noisy and noise-free environment. The reconstruction in this case is achieved with $10^{-4}$ precision. Note that the precision is lower compared with the noise-free case. When signal is corrupted with noise, simulations show that it is possible to reconstruct even heavily damaged signals. Increasing the number of iteration does not help much in lowering the MAE. Reconstruction can be affected mostly by adjusting μ and d parameters

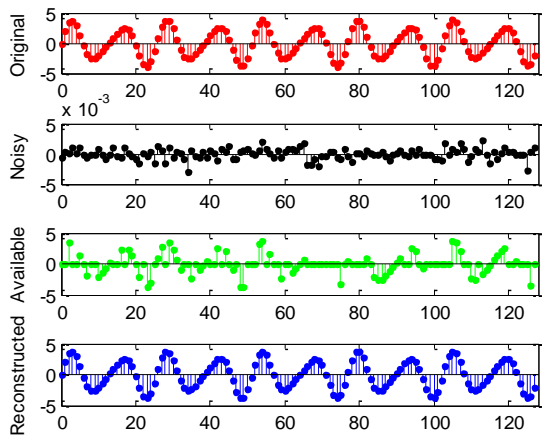

**Fig. 7. Original, Noisy, Available and Reconstructed signal**

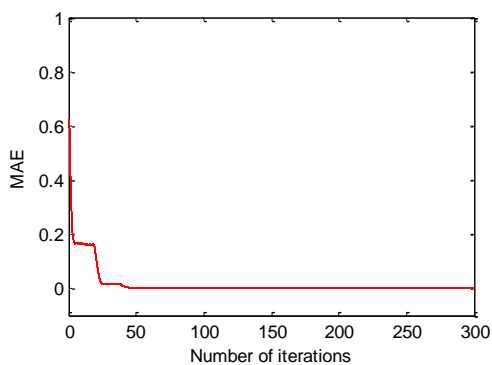

Fig. 8. The main absolute error for signal presented in Fig.7.

| Variance of noise | SNR (db) | MAE$_{MIN}$ 30 missing samples | MAE$_{MIN}$ 50 missing samples | MAE$_{MIN}$ 64 missing samples |
|---|---|---|---|---|
| 0.1 | 3 | 0.07774 | 0.07116 | 0.08731 |
| 0.15 | 4 | 0.09384 | 0.08952 | 0.0844 |
| 0.2 | 11 | 0.1231 | 0.1 | 0.1234 |
| 0.25 | 10 | 0.2059 | 0.162 | 0.1728 |
| 0.3 | 7 | 0.1975 | 0.1862 | 0.1817 |
| 0.35 | 14 | 0.2515 | 0.2188 | 0.1987 |
| 0.4 | 2 | 0.272 | 0.245 | 0.2204 |
| 0.45 | 8 | 0.3426 | 0.2925 | 0.331 |
| 0.5 | 13 | 0.3418 | 0.3274 | 0.3214 |

Table 2. MSE for different signal to noise ratios

In Table 2, we have presented the signal to noise ratio for different cases of missing samples (30, 50 and 64 (half) missing samples) and the precision achieved for each case. The other parameters are the same as in *Case 2*.

## IV. CONCLUSION

In this paper we have analyzed gradient-based algorithm for direct reconstruction of missing samples in a sparse signal. It is shown that algorithm gives satisfactory results even in the case when the signal is heavily damaged (more than 70 % of missing samples). Beside the signal with certain percent of missing samples, we have observed signals corrupted with external noise. Optimal measures were proposed for key parameters of the algorithm μ and *d*. Therefore, we can conclude that the most successful reconstruction is achieved for μ twice bigger than *d*.

## ACKNOWLEDGEMENT

The authors are thankful to Professors and assistants within the Laboratory for Multimedia Signals and Systems, at the University of Montenegro, for providing the ideas, codes, literature and results developed for the project CS-ICT (funded by the Montenegrin Ministry of Science).